\documentclass[conference]{IEEEtran}
\IEEEoverridecommandlockouts
\usepackage{subcaption}
\usepackage{cite}
\usepackage{amsmath,amssymb,amsfonts}
\usepackage{algorithmic}
\usepackage{multirow}
\usepackage{graphicx}
\usepackage{textcomp}
\usepackage{xcolor}
\usepackage{makecell}
\usepackage{booktabs}
\usepackage{multirow}
\def\BibTeX{{\rm B\kern-.05em{\sc i\kern-.025em b}\kern-.08em
    T\kern-.1667em\lower.7ex\hbox{E}\kern-.125emX}}

\makeatletter
\newcommand{\linebreakand}{%
  \end{@IEEEauthorhalign}
  \hfill\mbox{}\par
  \mbox{}\hfill\begin{@IEEEauthorhalign}
}

\makeatother

\begin{document}

\title{Rainbow Artifacts from Electromagnetic Signal Injection Attacks on Image Sensors
}



\author{
\IEEEauthorblockN{Youqian Zhang}
\IEEEauthorblockA{
\textit{The Hong Kong Polytechnic University}\\
you-qian.zhang@polyu.edu.hk
}
\and
\IEEEauthorblockN{Xinyu Ji}
\IEEEauthorblockA{\textit{The Hong Kong Polytechnic University}\\
csxji@comp.polyu.edu.hk
}
\linebreakand
\IEEEauthorblockN{Zhihao Wang}
\IEEEauthorblockA{\textit{The Hong Kong Polytechnic University}\\
zhi-hao-lucas.wang@connect.polyu.hk
}
\and
\IEEEauthorblockN{Qinhong Jiang}
\IEEEauthorblockA{\textit{The Hong Kong Polytechnic University}\\
qinhong.jiang@polyu.edu.hk
}
}

\maketitle

\begin{abstract}
Image sensors are integral to a wide range of safety- and security-critical systems, including surveillance infrastructure, autonomous vehicles, and industrial automation. 
These systems rely on the integrity of visual data to make decisions. 
In this work, we investigate a novel class of electromagnetic signal injection attacks that target the analog domain of image sensors, allowing adversaries to manipulate raw visual inputs without triggering conventional digital integrity checks.
We uncover a previously undocumented attack phenomenon on CMOS image sensors: rainbow-like color artifacts induced in images captured by image sensors through carefully tuned electromagnetic interference. 
We further evaluate the impact of these attacks on state-of-the-art object detection models, showing that the injected artifacts propagate through the image signal processing pipeline and lead to significant mispredictions. 
Our findings highlight a critical and underexplored vulnerability in the visual perception stack, highlighting the need for more robust defenses against physical-layer attacks in such systems.
\end{abstract}

\begin{IEEEkeywords}
Electromagnetic Interference, Image Sensor, Artificial Intelligence, Computer Vision, Object Detection
\end{IEEEkeywords}

\section{Introduction}

Modern intelligent systems have become integral to our daily lives, interacting with people to provide convenience, automation, and enhanced functionality. 
Among the many sensing components that enable these systems, image sensors play a pivotal role. 
They serve as the ``eyes'' of machines, allowing systems to perceive and interpret the physical environment.
The image sensors are deployed across a wide range of applications. 
For instance, surveillance cameras monitor public and private spaces for security threats, autonomous vehicles rely on them to detect and navigate obstacles on the road, and industrial robotic arms use them to perform precise assembly tasks. 
In all these scenarios, the accuracy and integrity of the images captured by the sensors are critical, as any distortion or manipulation of these images could lead to incorrect system behavior, potentially causing safety or security failures.

While image sensors are assumed to be trustworthy sources of visual data, recent studies~\cite{kohler2022signal, liu2025magshadow, ren2025ghostshot,dai2023magcode, jiang23glitchhiker, zhang2024esia, zhang2024modeling, liao2025your} have shown that they are vulnerable to ``Intentional Electromagnetic Signal Injection Attacks''. 
These attacks exploit the physical properties of electronic systems, specifically, the fact that cables and circuits can unintentionally act as antennas that pick up electromagnetic interference. 
By deliberately injecting crafted electromagnetic signals into the circuits, an attacker can superimpose malicious signals onto the original data being transmitted between the image sensor and its processing unit, corrupting the raw image data before being processed, without triggering any alarms in the system.
Previous studies have demonstrated that the attack can introduce visual artifacts,
which further propagate through the image processing pipeline and adversely affect downstream artificial intelligence (AI) tasks such as object detection, semantic segmentation, and face recognition. 
One surprising finding from prior work mentioned previously is that image signal processing (ISP), which is designed to clean up sensor noise and reconstruct high-quality images, is unable to mitigate the effects of such malicious interference.

Despite these important findings, a new and previously undocumented attack phenomenon has emerged from our investigation. 
We report a novel type of visual artifact introduced by carefully tuned electromagnetic signal injection so as to result in rainbow-like color strips that appear in images captured by Complementary Metal-Oxide-Semiconductor (CMOS) image sensors. 
This is especially significant because prior work has noted the difficulty of manipulating CMOS image sensors at the pixel level due to their readout mechanisms~\cite{kohler2022signal}. 
Our findings demonstrate that by precisely adjusting the electromagnetic signal parameters, it is indeed possible to inject structured, colorful interference into CMOS image data.
We further analyze how these rainbow-pattern attacks affect object detection models, quantifying the degradation in detection accuracy and consistency across multiple model architectures. 
Our results highlight the practical risks posed by such attacks, especially in safety- and security-critical systems that rely on machine vision.

The remainder of the paper is organized as follows. 
In Section~\ref{sec:background_and_related_work}, we review background knowledge and related work on electromagnetic signal injection attacks.
In Section~\ref{sec:methodology}, we present our methodology for generating rainbow-like artifacts.
Further, in Section~\ref{sec:object_detection}, we conduct experiments to evaluate the impacts of the attacks on various object detection models.
At last, we draw a conclusion in Section~\ref{sec:conclusion}.

\section{Background and Related Work}
\label{sec:background_and_related_work}

\subsection{Electromagnetic Signal Injection}

Electronic systems rely on internal communication between components, which is typically facilitated by conductive pathways such as cables or Printed Circuit Board (PCB) traces. 
These interconnects transmit both analog and digital signals essential to system functionality. 
However, a less desirable property of these conductors is their inherent ability to act as antennas, as mentioned earlier, unintentionally coupling with electromagnetic energy from their surroundings~\cite{paul2022introduction, wilson2010radiation}.

When exposed to external electromagnetic fields, these conductors can receive and propagate unwanted signals, similar to how a radio antenna picks up broadcast transmissions. 
This physical property introduces a critical attack surface: adversaries can exploit it by intentionally radiating electromagnetic signals toward the target device. 
Once intercepted by the cables or traces, these externally injected signals become superimposed on the legitimate signals being transmitted, potentially altering their waveform at the circuit level.

The success of such electromagnetic signal injection attacks generally depends on the frequency $f$ and power $A$ of the injection signal~\cite{yan2020sok}.
The injection frequency $f$ is a dominant parameter of an electromagnetic signal, which determines the coupling efficiency of the injection signal. Adversary can maximize the coupling efficiency by matching the injection frequency with the victim system's resonant frequency. The resonant frequency of the victim system can be identified through either theoretical analysis (mathematical modeling and simulation) or experimental testing (e.g., frequency sweeping test).
Studies~\cite{kune2013ghost,jiang2024ghosttype} have demonstrated that resonance-driven coupling can significantly increase the coupling efficiency of the injected signal, making the attacks more powerful. 
The power level $A$ of the injection signal determines the effectiveness of the attack. 
An injection signal with a higher power is more likely to dominate the legitimate signal, especially in analog domains: research has shown that increasing the transmitted power improves the amplitude of the coupled signal, thereby increasing the probability of successful interference or manipulation~\cite{zhang2022electromagnetic,jiang2024ghosttype}.

\subsection{Attack Impacts on Image Sensors}

\begin{figure}[t]
    \centering
    \begin{subfigure}[b]{0.32\linewidth}
        \includegraphics[width=\linewidth]{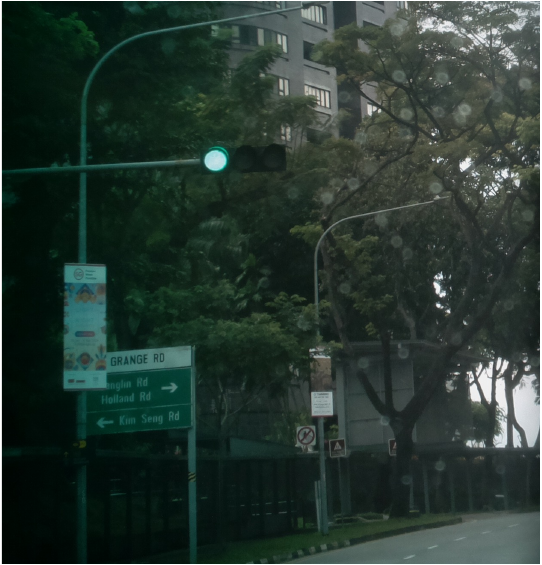}
        \caption{}
        \label{fig:street_original}
    \end{subfigure}
    \hfill
    \begin{subfigure}[b]{0.32\linewidth}
        \includegraphics[width=\linewidth]{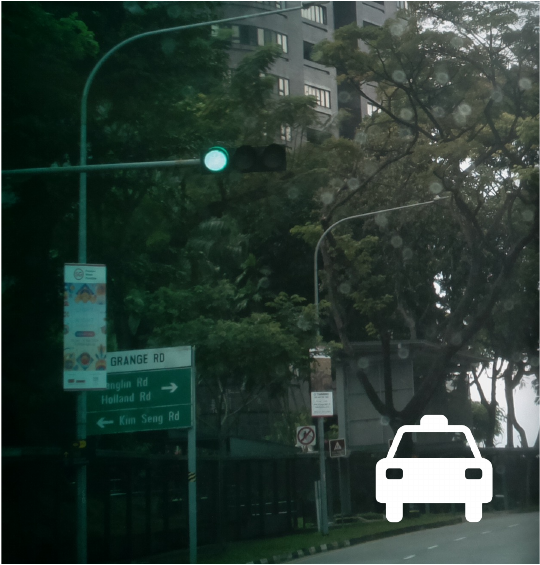}
        \caption{}
        \label{fig:street_ccd_attack}
    \end{subfigure}
    \hfill
    \begin{subfigure}[b]{0.32\linewidth}
        \includegraphics[width=\linewidth]{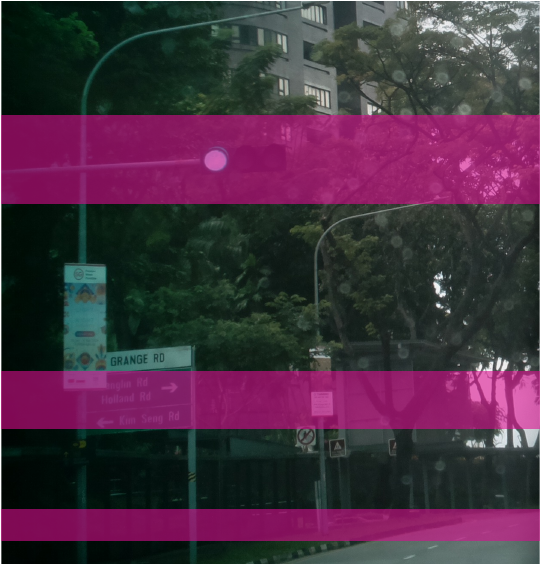}
        \caption{}
        \label{fig:street_cmos_attack}
    \end{subfigure}
    \caption{Comparison of the existing electromagnetic signal injection attacks on image sensors. (a) Ground truth image. (b) Attacks on a CCD sensor can introduce a false white car~\cite{kohler2022signal, liu2025magshadow, ren2025ghostshot}. (c) Attacks on a CMOS sensor can alter traffic light color, potentially misleading object recognition systems~\cite{jiang23glitchhiker, zhang2024esia, zhang2024modeling, liao2025your}. Note that the attack effects (b) and (c) are synthesized to illustrate their visual impacts.}
    \label{fig:previous_attack_impacts}
\end{figure}

Prior research on electromagnetic signal injection attacks has primarily focused on two widely used types of image sensors: Charge Coupled Device (CCD) and CMOS.
CCD sensors are known for their ability to produce high-resolution images and are still commonly used in scientific and industrial applications. 
One notable characteristic of CCD sensors is their vulnerability to fine-grained pixel-level manipulation. Prior work has demonstrated that adversaries can inject arbitrary patterns into images captured by CCD sensors through electromagnetic interference, effectively controlling individual pixels with high precision~\cite{kohler2022signal, liu2025magshadow, ren2025ghostshot}. 
This level of control enables a wide range of potential attacks, such as injecting fake objects or altering the appearance of real ones in the visual data.
In contrast, CMOS image sensors, which have become more prevalent due to their lower cost, are generally more resistant to such precise manipulation. 
However, recent studies~\cite{dai2023magcode, jiang23glitchhiker, zhang2024esia, zhang2024modeling, liao2025your} have shown that CMOS sensors are still vulnerable to electromagnetic signal injection. 
The attacks typically introduce visible artifacts, most notably purple stripes, into the captured images. 
Although less precise than attacks on CCDs, these artifacts can still significantly disrupt downstream computer vision tasks.

The impact of such attacks on AI-based perception systems can be severe. For instance, as shown in Figure~\ref{fig:previous_attack_impacts}, in object detection tasks, the presence of injected artifacts can cause the model to behave incorrectly in several ways. 
These include the creation effect, where the model detects objects that do not exist; the hiding effect, where existing objects go undetected; and the altering effect, where one object is misclassified as another~\cite{jiang23glitchhiker}. 
These misbehaviors are especially dangerous in safety-critical applications. 
A striking example is in autonomous driving, where an attacker could potentially obscure a red traffic light in the image, leading the vehicle to misinterpret the scene and fail to stop, with potentially catastrophic consequences. 

\section{Methodology}
\label{sec:methodology}

\subsection{Experiment Setup}

\begin{figure}[t]
    \centering
    \includegraphics[width=0.9\linewidth]{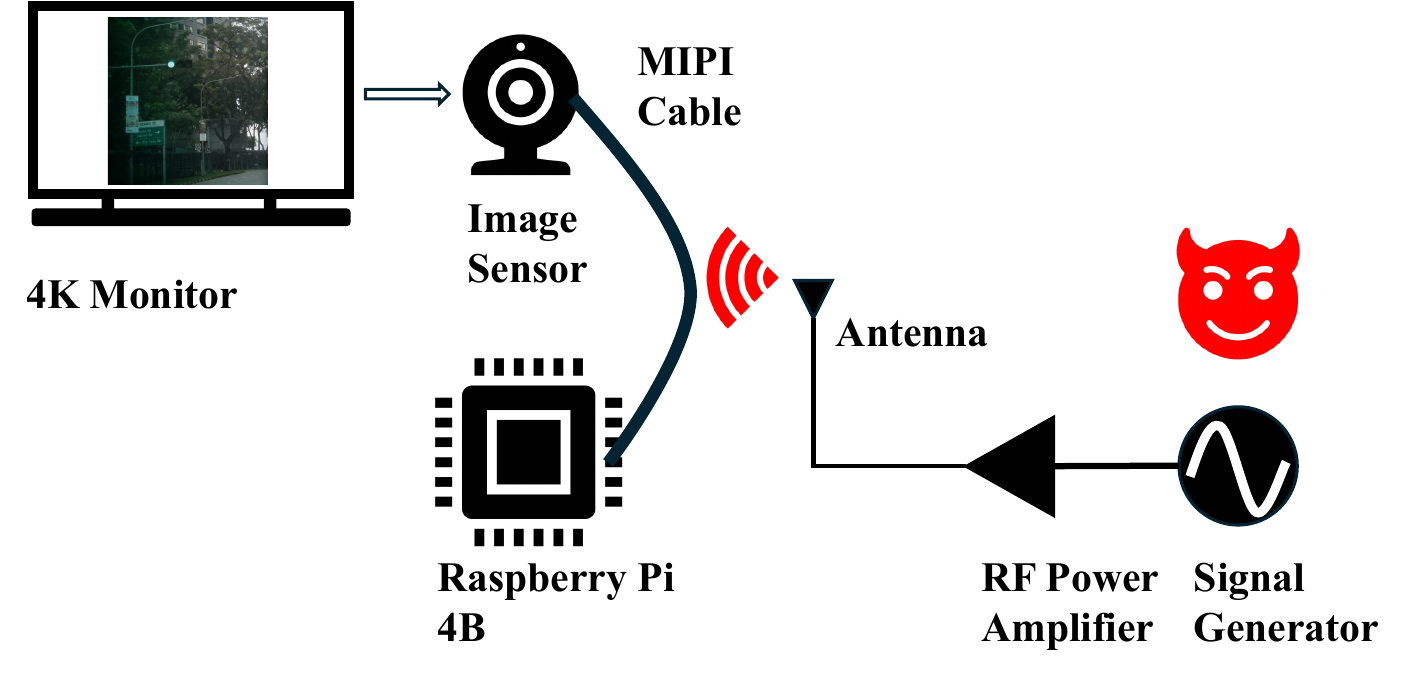}
    \caption{Experimental setup for electromagnetic signal injection.}
    \label{fig:attack-setup}
\end{figure}

To investigate the feasibility and effects of electromagnetic signal injection attacks on image sensors, we set up a controlled experimental environment using commercially available hardware components, as depicted in Figure~\ref{fig:attack-setup}.
Our experimental platform consists of a Raspberry Pi 4B connected to an OV5647 CMOS image sensor module via a standard MIPI CSI-2 cable. The OV5647 is a widely used 5-megapixel sensor, commonly found in embedded vision applications. To simulate realistic visual input, a 4K display was used to display street scenes\footnote{An Institutional Review Board (IRB) approval has been obtained to allow the researchers to capture photos of public places and use them for this research.}, which the image sensor captured in real time.

We use a signal generator to produce an amplitude-modulated (AM) continuous sine wave, with the carrier sweeping from 20 MHz to 160 MHz (the generator's operational limits) and the baseband signal ranging from 0 kHz to 50 kHz.

To ensure the injected signal is strong enough to couple into the target system, the modulated signal is passed through a 3-watt RF power amplifier. 
The amplified signal is then transmitted through an omnidirectional antenna, which is positioned approximately 5 centimeters away from the cable connecting the image sensor to the Raspberry Pi. 
This close proximity ensures efficient electromagnetic coupling between the antenna and the cable, leveraging the cable’s unintentional antenna-like behavior.

\subsection{Observed Phenomenon}

\begin{figure}[t]
    \centering
    \begin{subfigure}[b]{0.57\linewidth}
        \includegraphics[width=\linewidth]{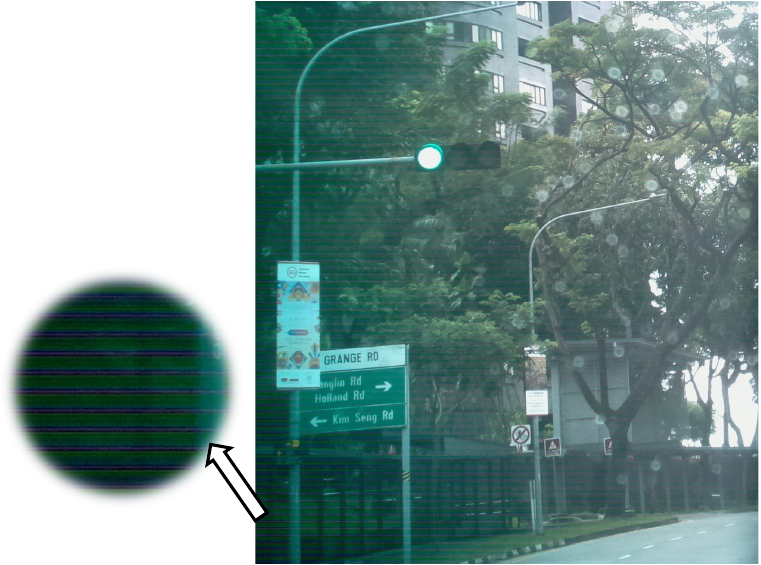}
        \caption{}
        \label{fig:rainbow-dense}
    \end{subfigure}
    \hfill
    \begin{subfigure}[b]{0.38\linewidth}
        \includegraphics[width=\linewidth]{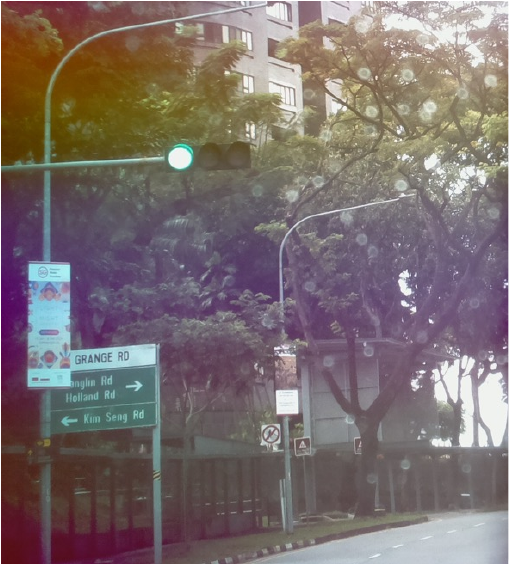}
        \caption{}
        \label{fig:rainbow-sparse}
    \end{subfigure}
    \caption{Illustrations of (a) dense rainbow artifacts generated at 26.16kHz and (b) sparse rainbow artifacts generated at 46.16kHz}
    \label{fig:rainbow-comparison}
\end{figure}

During our experiments, we identified a distinct and previously undocumented visual artifact, a rainbow-like color pattern, that emerges only under specific electromagnetic signal conditions. 
This effect was reproducibly observed when the carrier frequency of the injected signal was precisely set to 84.68~MHz, and the frequency of the baseband signal was swept within the range of 26.16~kHz to 46.16~kHz. 
Under these conditions, the image sensor captured a series of semi-transparent, colored bands, primarily consisting of red, green, and blue hues, which blend together to resemble natural rainbow patterns, as shown in Figure~\ref{fig:rainbow-comparison}. 
These artifacts appeared superimposed over the original scene without completely obscuring the underlying content, giving the effect a translucent quality.

Notably, the baseband frequency had a direct influence on the density of the rainbow bands. 
When the baseband frequency was higher within the specified range, the resulting patterns became more sparse (Figure.~\ref{fig:rainbow-sparse}), with wider spacing between the colored bands. Conversely, lower baseband frequencies resulted in denser patterns, with the colored stripes appearing closer together (Figure.~\ref{fig:rainbow-dense}). This phenomenon suggests a strong interaction between the modulation dynamics and the image sensor’s readout mechanism, particularly in how the injected signal interferes with the analog signal processing stage before digitization.

This attack differs from the previous one that caused purple stripes in that it does not result in packet loss (packet loss is a necessary cause of the purple artifacts). 
Instead, it demonstrates control over pixel colors. Based on the observed effects, it is highly likely that the attack signal directly affects the pixels of the image sensor.


\section{Impacts on Object Detection}
\label{sec:object_detection}

\begin{table*}[!htbp]
\centering
\vspace{0.3cm}
\caption{Performance of electromagnetic signal injection attacks on object detection models.}
\label{tab:result}
\resizebox{\textwidth}{!}{%
\begin{tabular}{
  l      
  l      
  l      
  >{\centering}p{1.7cm} 
  >{\centering}p{1.7cm} 
  >{\centering}p{1.7cm} 
  >{\centering\arraybackslash}p{1.7cm} 
}
\toprule
\textbf{Model} &
\textbf{Version} &
\textbf{Scenario} &
\textbf{Precision} &
\textbf{Recall} &
\textbf{mAP@0.5} &
\textbf{mAP@0.5:0.95} \\
\midrule

\multirow{8}{*}{YOLO} 
  & yolo5nu~\cite{yolov5}  & w/o Attack & 0.753 & 0.719 & 0.775 & 0.606 \\
  &                        & Attack     & 0.628 & 0.513 & 0.632 & 0.470 \\
  & yolo11n~\cite{yolov11} & w/o Attack & 0.784 & 0.709 & 0.774 & 0.618 \\
  &                        & Attack     & 0.678 & 0.550 & 0.647 & 0.509 \\
  & yolo10n~\cite{yolov10} & w/o Attack & 0.745 & 0.774 & 0.820 & 0.651 \\
  &                        & Attack     & 0.667 & 0.607 & 0.672 & 0.521 \\
  & yolo8n~\cite{yolov8}   & w/o Attack & 0.736 & 0.750 & 0.797 & 0.698 \\
  &                        & Attack     & 0.810 & 0.555 & 0.684 & 0.540 \\
\midrule

\multirow{2}{*}{Faster-RCNN~\cite{fasterrcnn}} 
  & faster-rcnn\_x101-64x4d\_fpn\_ms-3x\_coco & w/o Attack & 0.501 & 0.742 & 0.832 & 0.608 \\
  &                                           & Attack     & 0.492 & 0.735 & 0.793 & 0.544 \\
\midrule

\multirow{2}{*}{DETR~\cite{detr}}
  & detr\_r50\_8xb2-150e\_coco & w/o Attack & 0.307 & 0.745 & 0.724 & 0.430 \\
  &                            & Attack     & 0.284 & 0.702 & 0.657 & 0.377 \\
\bottomrule
\end{tabular}
}
\end{table*}

\subsection{Data Collection}

We collected data using the experimental setup illustrated in Figure~\ref{fig:attack-setup}. A total of 93 unique scenes were captured, covering both attack and no-attack conditions. Each scene includes real-world urban environments displayed on a 4K screen and recorded by the image sensor. For each captured image, we performed manual annotation of key object categories, including person, car, motorcycle, bus, truck, and traffic light. These annotations serve as ground truth for evaluating the impact of electromagnetic signal injection on object detection performance.

\subsection{Models and Metrics}
We selected six object detectors from three main categories: four one-stage object detectors (e.g., YOLO, including its variants such as yolo5nu, yolo11n, yolo10n, and yolo8n)~\cite{yolov5,yolov10,yolov11,yolov8}, one two-stage object detector (e.g., fasterRCNN)~\cite{fasterrcnn}, and one transformer-based object detector (e.g., DETR)~\cite{detr}.

To evaluate the impact of electromagnetic signal injection on object detection performance, we adopt three widely used metrics: Precision, Recall, and mean Average Precision (mAP). 
Precision measures the proportion of correctly predicted positive detections out of all positive predictions, and is defined as: $\text{Precision} = \frac{TP}{TP + FP}$, where $TP$ denotes the number of true positives and $FP$ denotes the number of false positives.
Recall quantifies the proportion of correctly predicted positive detections out of all actual positive instances: $\text{Recall} = \frac{TP}{TP + FN}$, where $FN$ is the number of false negatives.
Furthermore, following the COCO evaluation protocol~\cite{lin2014microsoft}, we report:
\begin{itemize}
    \item mAP@0.5: the mean AP calculated at a fixed Intersection over Union (IoU) threshold of 0.5.
    \item mAP@0.5:0.95: the mean AP averaged over IoU thresholds from 0.5 to 0.95 in steps of 0.05.
\end{itemize}

\begin{figure}[t]
    \centering
    \begin{subfigure}[b]{0.32\linewidth}
        \includegraphics[width=\linewidth]{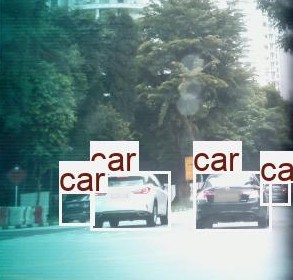}
        \caption{Ground truth}
        \label{fig:Ground-truth}
    \end{subfigure}
    \hfill
    \begin{subfigure}[b]{0.32\linewidth}
        \includegraphics[width=\linewidth]{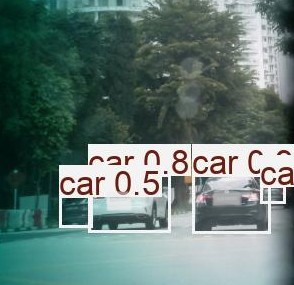}
        \caption{w/o Aattack}
        \label{fig:w/o Aattack}
    \end{subfigure}
    \hfill
    \begin{subfigure}[b]{0.32\linewidth}
        \includegraphics[width=\linewidth]{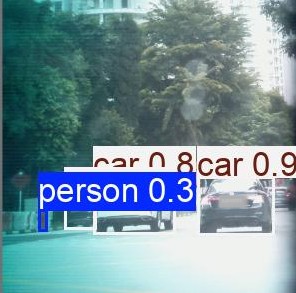}
        \caption{Under Attack}
        \label{fig:attack}
    \end{subfigure}
    \caption{The effects of electromagnetic signal injection attacks on object detection, where the ground truth refers to the manually annotated labels within the dataset.}
    \label{fig:example}
\end{figure}

\subsection{Results and Discussion}

The results are summarized in Table~\ref{tab:result}, and it can be found that there is a clear degradation in detection performance across all models when subjected to electromagnetic signal injection attacks. For the YOLO series, the yolo5nu model shows a drop in precision from 0.753 to 0.628 and in recall from 0.719 to 0.513. mAP@0.5 decreases from 0.775 to 0.632, and mAP@0.5:0.95 drops from 0.606 to 0.470. Similar trends are observed for yolo11n, yolo10n, and yolo8n. In particular, the Recall and mAP metrics experience the most significant reductions, indicating a substantial loss in the models’ ability to correctly detect objects.

Faster-RCNN also exhibits performance degradation under attack. Although the changes in Precision and Recall are less pronounced, the decrease in mAP highlights a reduction in overall detection accuracy. Furthermore, the DETR model is affected similarly. These results confirm that electromagnetic signal injection attacks negatively impact detection performance across different model architectures.

The most affected metrics, as just mentioned, are Recall and mAP, which are critical for reliable object recognition in safety-critical applications such as autonomous driving. 
Specifically, regarding the recall, its degradation increases the risk of not detecting obstacles, which can lead to safety incidents. 
For instance, as shown in Fig~\ref{fig:example}, the attack causes the object detection model to affect confidence in some key obstacles, resulting in failure to detect a car, which may cause the autonomous vehicle to take unsafe actions, such as accelerating dangerously near other road users.
Moreover, the model might incorrectly detect a pedestrian or trigger abrupt emergency braking, further endangering passengers and nearby traffic participants.


\section{Conclusion and Future Work}
\label{sec:conclusion}

In this work, we demonstrate that by carefully selecting the carrier and baseband frequencies, we are able to produce a distinct ``rainbow'' effect that appears as semi-transparent red, green, and blue bands across the captured scene. 
Our experiments show that such physical-layer interference can significantly degrade object detection performance, with notable impacts on precision, recall, and mean Average Precision (mAP). 
In particular, safety-critical objects such as those in traffic scenarios are shown to be susceptible to attack-induced misdetections, raising concerns about the robustness of vision-based perception systems in real-world deployments.
In future work, we plan to explore a broader range of attack configurations, including different modulation schemes, frequencies, and power levels, to further understand the boundary conditions under which such attacks succeed. 
Additionally, we aim to investigate the underlying physical and electronic mechanisms that manifest as structured visual patterns. 
Understanding the root cause of these effects may enable the development of more effective countermeasures that can resist or adapt to such adversarial physical-world perturbations.

\bibliographystyle{ieeetr}
\bibliography{main}

\begin{thebibliography}{10}

\bibitem{kohler2022signal}
S.~K{\"o}hler, R.~Baker, and I.~Martinovic, ``{Signal Injection Attacks against CCD Image Sensors},'' in {\em Proc. 2022 ACM ASIA Conference on Computer and Communications Security (ACM ASIACCS 2022)}, ACM, 2022.

\bibitem{liu2025magshadow}
Z.~Liu, F.~Lin, Z.~Ba, L.~Lu, and K.~Ren, ``{MagShadow: Physical Adversarial Example Attacks via Electromagnetic Injection},'' {\em IEEE Transactions on Dependable and Secure Computing}, 2025.

\bibitem{ren2025ghostshot}
Y.~Ren, Q.~Jiang, C.~Yan, X.~Ji, and W.~Xu, ``{GhostShot: Manipulating the Image of CCD Cameras with Electromagnetic Interference},'' in {\em NDSS}, 2025.

\bibitem{dai2023magcode}
D.~Dai, Z.~An, Q.~Pan, and L.~Yang, ``{Magcode: NFC-enabled Barcodes for NFC-disabled Smartphones},'' in {\em Proceedings of the 29th Annual International Conference on Mobile Computing and Networking}, pp.~1--14, 2023.

\bibitem{jiang23glitchhiker}
Q.~Jiang, X.~Ji, C.~Yan, Z.~Xie, H.~Lou, and W.~Xu, ``{GlitchHiker: Uncovering Vulnerabilities of Image Signal Transmission with IEMI},'' in {\em The 32nd USENIX Security Symposium}, 2023.

\bibitem{zhang2024esia}
Y.~Zhang, C.~Yang, Y.~Fu, Q.~Jiang, C.~Yan, S.-Y. Chau, G.~Ngai, H.-v. Leong, X.~Luo, and W.~Xu, ``{Understanding Impacts of Electromagnetic Signal Injection Attacks on Object Detection},'' in {\em IEEE International Conference on Multimedia and Expo}, IEEE, 2024.

\bibitem{zhang2024modeling}
Y.~Zhang, M.~Cheung, C.~Yang, X.~Zhai, Z.~Shen, X.~Ji, E.~Y. Fu, S.-Y. Chau, and X.~Luo, ``{Modeling Electromagnetic Signal Injection Attacks on Camera-based Smart Systems: Applications and Mitigation},'' {\em arXiv preprint arXiv:2408.05124}, 2024.

\bibitem{liao2025your}
W.~Liao, S.~Yan, Y.~Zhang, X.~Zhai, Y.~Wang, and E.~Y. Fu, ``{Is Your Autonomous Vehicle Safe? Understanding the Threat of Electromagnetic Signal Injection Attacks on Traffic Scene Perception},'' {\em Proceedings of the AAAI Conference on Artificial Intelligence}, 2025.

\bibitem{paul2022introduction}
C.~R. Paul, R.~C. Scully, and M.~A. Steffka, {\em {Introduction to Electromagnetic Compatibility}}.
\newblock John Wiley \& Sons, 2022.

\bibitem{wilson2010radiation}
P.~F. Wilson, ``{Radiation Patterns of Unintentional Antennas: Estimates, Simulations, and Measurements},'' in {\em 2010 Asia-Pacific International Symposium on Electromagnetic Compatibility}, pp.~985--989, IEEE, 2010.

\bibitem{yan2020sok}
C.~Yan, H.~Shin, C.~Bolton, W.~Xu, Y.~Kim, and K.~Fu, ``{SoK: A Minimalist Approach to Formalizing Analog Sensor Security},'' in {\em 2020 IEEE Symposium on Security and Privacy (SP)}, pp.~233--248, IEEE, 2020.

\bibitem{kune2013ghost}
D.~F. Kune, J.~Backes, S.~S. Clark, D.~Kramer, M.~Reynolds, K.~Fu, Y.~Kim, and W.~Xu, ``{Ghost Talk: Mitigating EMI Signal Injection Attacks against Analog Sensors},'' in {\em 2013 IEEE Symposium on Security and Privacy}, pp.~145--159, IEEE, 2013.

\bibitem{jiang2024ghosttype}
Q.~Jiang, Y.~Ren, Y.~Long, C.~Yan, Y.~Sun, X.~Ji, K.~Fu, and W.~Xu, ``Ghosttype: The limits of using contactless electromagnetic interference to inject phantom keys into analog circuits of keyboards,'' in {\em Network and Distributed Systems Security (NDSS) Symposium}, 2024.

\bibitem{zhang2022electromagnetic}
Y.~Zhang, {\em {Electromagnetic Signal Injection Attacks on Embedded Systems: Modeling and Detection}}.
\newblock PhD thesis, University of Oxford, 2022.

\bibitem{yolov5}
G.~Jocher, ``Ultralytics yolov5,'' 2020.

\bibitem{yolov11}
G.~Jocher and J.~Qiu, ``Ultralytics yolo11,'' 2024.

\bibitem{yolov10}
e.~a. Ao~Wang, ``Yolov10: Real-time end-to-end object detection,'' {\em arXiv preprint arXiv:2405.14458}, 2024.

\bibitem{yolov8}
G.~Jocher, A.~Chaurasia, and J.~Qiu, ``Ultralytics yolov8,'' 2023.

\bibitem{fasterrcnn}
S.~Ren, K.~He, R.~Girshick, and J.~Sun, ``Faster r-cnn: Towards real-time object detection with region proposal networks,'' {\em IEEE Transactions on Pattern Analysis and Machine Intelligence}, Jun 2017.

\bibitem{detr}
N.~Carion, F.~Massa, G.~Synnaeve, N.~Usunier, A.~Kirillov, and S.~Zagoruyko, ``End-to-end object detection with transformers,'' in {\em ECCV}, 2020.

\bibitem{lin2014microsoft}
T.-Y. Lin, M.~Maire, S.~Belongie, J.~Hays, P.~Perona, D.~Ramanan, P.~Doll{\'a}r, and C.~L. Zitnick, ``Microsoft coco: Common objects in context,'' in {\em Computer vision--ECCV 2014: 13th European conference, zurich, Switzerland, September 6-12, 2014, proceedings, part v 13}, pp.~740--755, Springer, 2014.

\end{thebibliography}

\end{document}